\documentclass[prb, twocolumn]{revtex4}

\usepackage{float}
\usepackage[caption = false]{subfig}

\usepackage{amsmath}
\usepackage{amsfonts}

\usepackage{epsfig} 

\usepackage{epstopdf}
\newcommand{\Ket}[1]{\left|#1  \right>}

\newcommand{\Braket}[1]{\left<#1  \right>}

\def\begeq{\begin{equation}}
\def\endeq{\end{equation}}

\def\begeqar{\begin{eqnarray}}
\def\endeqar{\end{eqnarray}}

\usepackage{color}

\begin{document}

\title{Multifractal Orthogonality Catastrophe in 1D Random Quantum Critical Points}

\author{Romain Vasseur and Joel E. Moore}

\affiliation{Department of Physics, University of California, Berkeley, Berkeley CA 94720, USA}
\affiliation{Materials Science Division, Lawrence Berkeley National Laboratory, Berkeley CA 94720, USA}
\date{\today}

\begin{abstract}

We study the response of random singlet quantum critical points to local perturbations. Despite being insulating, these systems are dramatically affected by a local cut in the system, so that the overlap $G=\left|\Braket{\Psi_B |\Psi_A}\right|$ of the groundstate wave functions with and without a cut vanishes algebraically in the thermodynamic limit. We analyze this Anderson orthogonality catastrophe in detail using a real-space renormalization group approach. We show that both the typical value of the overlap G and the disorder average of $G^\alpha$ with $\alpha>0$ decay as power-laws of the system size. In particular, the disorder average of $G^\alpha$ shows a ``multifractal'' behavior, with a non-trivial limit $\alpha \to \infty$ that is dominated by rare events. We also discuss the case of more generic local perturbations and generalize these results to local quantum quenches.

\end{abstract}

\maketitle

\section{Introduction}

One of the most basic questions about quantum dynamics is the response of a system to a sudden change of the Hamiltonian in a small region in space.  This local quantum quench can be expected to generate a superposition of ground and excited states, even if the system was prepared in its ground state before the quench, and the structure of this superposition is quite complicated in many-particle systems.  An early experimental example of a local quantum quench from solid-state physics is the excitation of a core level by an incident photon of high energy, which appears to the conduction electrons as a sudden change in the local electrostatic potential.  The response of the Fermi sea to this change is quite interesting~\cite{PhysRev.163.612,PhysRev.178.1097} and led to the development of important concepts in many-body theory such as that of an orthogonality catastrophe~\cite{PhysRevLett.18.1049}, which is a decay with system size in the overlap of ground states of the initial and final Hamiltonian.

The orthogonality catastrophe appears in the vacuum-to-vacuum transition probability (also known as ground state fidelity or ground state overlap): what is the probability that the system after the quench is in the ground state of the new Hamiltonian?  In metallic systems, this leads to an overlap that decays as a power-law in system size $L$,
\begin{equation}
G=|\Braket{\Psi_B |\Psi_A}| \propto L^{-\alpha}.
\end{equation}
In words, the orthogonality catastrophe in a metal is that the ground states before and after a local potential change are actually orthogonal in the thermodynamic limit.  In experiments, the orthogonality catastrophe is essential for understanding the frequency dependence of optical absorption near the X-ray edge where such absorption by a core level becomes energetically allowed, and is also closely related to non linear $I-V$ characteristics in quantum dots, or the Kondo effect~\cite{9780511470752} in magnetic alloys.  The main goal of this paper is to study the orthogonality catastrophe that emerges in  strongly disordered quantum systems in one dimension.

In recent years, quantum quenches in translation-invariant one-dimensional systems have been studied analytically with considerable success as powerful methods from conformal field theory~\cite{Calabrese:2006,1742-5468-2007-10-P10004, PhysRevLett.109.260601}, integrability~\cite{1367-2630-12-5-055015, PhysRevLett.106.227203, PhysRevLett.110.257203}, and numerics~\cite{PhysRevLett.98.180601,PhysRevLett.98.210405} are available.  Systems with random impurities are in general more complicated but a rare example of a strongly disordered interacting system that can be studied analytically is the random-singlet quantum critical point~\cite{PhysRevB.22.1305,FisherRSRG2}, whose disorder-averaged equal-time correlation functions scale as simple power-laws but whose dynamical properties are relatively complicated~\cite{PhysRevLett.84.3434}.  Two unexpected findings of the present work are that there is a power-law orthogonality catastrophe similar to that in metals, even though the system is insulating, and that the orthogonality catastrophe is ``multifractal'' in the sense that different powers of the disorder-averaged overlap $G$ scale with nontrivially different powers of system size.  

The main quench we discuss in detail is a ``cut'' in the system that disconnects the left and right halves.  This is a physical cut, {\it i.e.}, a change in the Hamiltonian, not the mathematical division of the Hilbert space into two parts that is used to calculate entanglement.  However, there are connections between the orthogonality catastrophe and the disorder-averaged entanglement properties of the random singlet phase, which have been an active subject~\cite{refaelreview}. The crucial point is that $G$ keeps track of how many singlets are affected by the cut in the system -- a quantity different from entanglement whose statistics can be accessed using real space renormalization group techniques. The result that the orthogonality catastrophe at the random-singlet critical point is similar to that in a metal, but that there is a difference in the multifractal properties, are perhaps believable in light of the fact that entanglement entropy is known to scale similarly to that of a metal (or other conformally invariant system), while the entanglement spectrum behaves differently~\cite{PhysRevB.83.045110}.

The remainder of this Introduction reviews briefly some relevant recent progress on related questions.  New tools for quantum quenches include, on the experimental side, atomic systems where the absence of phonons and resulting long decoherence times lead to many ways to generate quantum quenches and observe them while the dynamics remain quantum-mechanical.  On the theoretical side, progress in time-dependent density-matrix renormalization group methods~\cite{white,schollwoeck} allows simulation of systems that are neither integrable nor conformally invariant and hence difficult to study analytically.  Strongly disordered interacting systems remain challenging for numerics but are of great current importance, partly because the same real-space renormalization-group (RSRG) methods we apply have been important for recent progress in many-body localization (MBL)~\cite{VoskAltmanPRL13,PekkerRSRGX, PhysRevLett.112.217204,HeisenbergRSRGX}, the existence of localized behavior at nonzero temperature or energy density (see Refs.~\onlinecite{RevMBL,ReviewMBLEhud} for recent reviews).  While the ground-state fidelity is quite a different property than MBL, it is hoped that some of the technical developments here will prove useful.

The multifractal properties of one-particle wavefunctions in disordered systems have been studied for many years~\cite{Mirlin2000259,RevModPhys.80.1355}, including recent progress on new sets of exponents appearing at edges and corners~\cite{Subramaniam2006}.  This problem is loosely connected to the random-singlet phase of spin chains in the following way.  Essentially the same random-singlet phase appears both for the interacting Heisenberg/XXX chain as for the XX spin chain, but only the latter can be mapped via the Jordan-Wigner transformation onto a non-interacting one-dimensional electron hopping system with particle-hole symmetry in the energy spectrum resulting from sublattice symmetry.  Using this trick for the XX case, the orthogonality catastrophe we find can be reformulated as a statement about a different kind of multifractality that appears in the comparison of one-electron wavefunctions between two locally different critical Hamiltonians, and in that representation is closer to Anderson's original study.  We wish to point out that the RSRG calculations we report are expected to be valid even for interacting systems in the random-singlet phase, although the ability to check them against microscopic numerics is limited to the XX case, and even then rather nontrivial.

The remainder of this paper is organized as follows.  Section~\ref{sec1} introduces the key features of the random-singlet phase and the RSRG techniques used in Section~\ref{sec2} to calculate the ground-state overlap.  Section~\ref{sec3} discusses the multifractal spectrum of the orthogonality catastrophe using a generating-function approach.
Section~\ref{sec4} presents numerical results on the main quantities of interest, and Section~\ref{sec5} discusses which aspects of our analysis are expected to be general to other perturbations and to practical quantum quenches.

\section{Groundstate overlap and strong disorder renormalization group}
\label{sec1}

\subsection{Local perturbations and overlaps}

Groundstate overlaps, or fidelities, are a clear-cut way to characterize phases of matter. They are particularly useful to probe the response of a system to local perturbations: denoting by $\Ket{\Psi_A}$ and $\Ket{\Psi_B}$ the groundstate wave functions without and with a local perturbation, respectively, how does the overlap (fidelity) $G=\left|\Braket{\Psi_B |\Psi_A}\right|$ vary as a function of system size $L$ (considering for simplicity a 1D system)? This question goes back to the idea of Anderson orthogonality catastrophe~\cite{PhysRevLett.18.1049}, and in the case where the local perturbation corresponds to a cut in the system, is somewhat related to the  bipartite entanglement entropy, while at the same time being fundamentally different and much simpler conceptually. Studying  the response of a system to a local perturbation is also very natural from the point of view of local quantum quenches: starting at time $t=0$ from the groundstate $\Ket{\Psi_A}$ of the system without perturbation and time-evolving with the Hamiltonian $H_B$ of the system with the local perturbation, how does the time-dependent overlap $G(t)=\Braket{\Psi_A |{\rm e}^{i H_A t} {\rm e}^{-i H_B t}|\Psi_A}$ (known as the Loschmidt echo in the literature) behave at long times? In this dynamical setting, this problem is closely related to the X-ray edge singularity problem~\cite{PhysRev.163.612,PhysRev.178.1097}, and the Loschmidt echo can be related to optical absorption spectra~\cite{PhysRevLett.106.107402,Latta:2011aa,PhysRevLett.108.190601,PhysRevLett.110.230601}.  For some recent physical applications of the Loschmidt echo to cold atoms or condensed matter systems, see Refs.~\onlinecite{PhysRevX.2.041020,PhysRevLett.111.046402, PhysRevLett.110.240601, PhysRevLett.112.246401,PhysRevLett.112.146804,PhysRevX.4.041007}.

   \begin{figure}[t!]
\includegraphics[width=1.0\columnwidth]{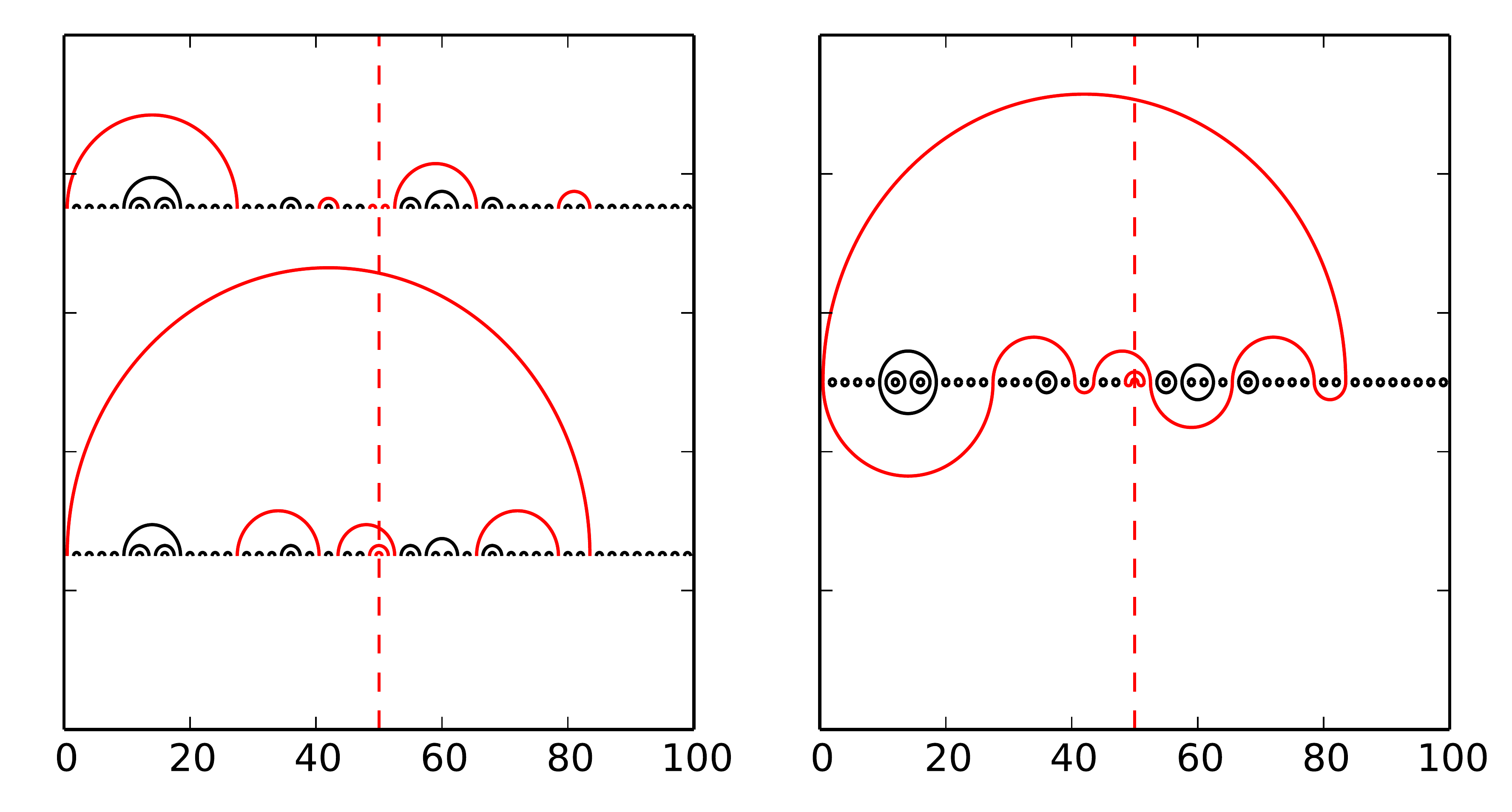}
\caption{Left:  Example of random singlet configurations on $L=100$ sites with ($\Ket{\Psi_B}$, top) and without ($\Ket{\Psi_A}$, bottom) a cut in the middle of the system (dashed vertical line), for the same disorder configuration. Only the red singlets are affected by the cut. Right: The overlap $G=\Braket{\Psi_B |\Psi_A}$ is computed by counting the number of loops formed by glueing these configurations together. In this example, we find $G=\left( \frac{1}{2}\right)^{4}$ consistent with $n_e=4$ singlets crossing the cut in $\Ket{\Psi_A}$ and $n=6$ singlets modified by the cut (see eq~\eqref{eqGnne}). }
\label{figureConfig}
\end{figure}

We will mostly focus on the case of a cut in the system, and study the ``static'' response $G=\left| \Braket{\Psi_B |\Psi_A} \right|$ as a function of system size $L$ -- we will also discuss the case of more generic local perturbations and come back to local quantum quenches in Sec.~\ref{secQuenches}. We will ignore all phase factors and  implicitly focus on the magnitude of the overlap $G=\left| \Braket{\Psi_B |\Psi_A} \right|=\Braket{\Psi_B |\Psi_A}$. To be more precise, let $\Ket{\Psi_A}$ be the groundstate of a system of size $L=2N$, and $\Ket{\Psi_B}$ be the groundstate of the same system cut in two halves $[0,N]$ and $[N,L=2N]$. For gapped ground states, the overlap $G=\Braket{\Psi_B |\Psi_A}$ remains finite in the thermodynamic limit. For gapless systems however, $G$ vanishes as a power-law as $L \to \infty$, with some exponent related to the central charge of the corresponding conformal field theory~\cite{1742-5468-2011-03-L03002}. For strongly disordered systems in 1D, the average of $G$ is non-vanishing because of Anderson localization~\cite{PhysRevB.65.081106} -- although it was pointed out recently that localized systems do suffer from a ``statistical'' version of the orthogonality catastrophe~\cite{2014arXiv1411.2616K}. In this paper, we will be interested in random singlet quantum critical points, that are examples of infinite randomness fixed points. Despite being insulating, they exhibit critical properties such as algebraically decaying averaged correlation functions~\cite{FisherRSRG1,FisherRSRG2,PhysRevB.51.6411}, logarithmic scaling of entanglement~\cite{RefaelMoore}, and energy-length scaling~\cite{FisherRSRG2,PhysRevB.51.6411} 
\begin{equation}
\ln \frac{1}{E} \sim L^\psi,
\label{eqScalingtE}
\end{equation}
(with $\psi=1/2$ for the examples treated in this paper) instead of the usual quantum-critical relation $E\sim L^{-z}$.

For concreteness, we will focus on the antiferromagnetic spin-$\frac{1}{2}$ random-bond Heisenberg chain
\begin{equation}
H=\sum_i J_i \vec{S}_i . \vec{S}_{i+1}.
\label{eqXXX}
\end{equation}
More generally, our results will also apply directly to the anisotropic XXZ chain~\cite{FisherRSRG2}
\begin{equation}
H=\sum_i J_i \left( S^x_i S^x_{i+1} + S^y_i S^y_{i+1} + \Delta S^z_i S^z_{i+1}\right),
\label{eqXXZ}
\end{equation}
for $-1/2<\Delta<1$, and we will discuss generalizations to other random systems (namely, the transverse field Ising chain~\cite{FisherRSRG1} and anyonic chains~\cite{PhysRevLett.99.140405}) in Sec.~\ref{subsecTyp}.

\subsection{Real space renormalization group}

Disorder is a relevant perturbation (in the renormalization group sense) to the pure spin-$\frac{1}{2}$ Heisenberg chain. At low-energy, the disorder strength flows to infinity and the universal properties of the random bond Heisenberg chain can be accessed using a real space renormalization group (RSRG) approach~\cite{FisherRSRG2}, valid at strong randomness. The key idea of this approach is to focus on the strongest bond of the chain $J_i$. Assuming strong disorder, this bond is typically much larger than its neighbors $J_i \gg J_{i-1}, J_{i+1}$, so to leading order we can diagonalize this strong bond and form a singlet between the spins $S_i$ and $S_{i+1}$, and then deal with the rest of the chain perturbatively. Virtual fluctuations induce an effective Heisenberg coupling between the spins $S_{i-1}$ and $S_{i+2}$ given by~\cite{PhysRevLett.43.1434,PhysRevB.22.1305}
\begin{equation}
J_{\rm eff} = \frac{J_{i-1} J_{i+1}}{2 J_i},\label{eqdecimationrule}
\end{equation}
with $J_{\rm eff} \ll J_{i-1},  J_i,  J_{i+1}$ at strong disorder. Repeating this process produces singlets at increasingly long length scales, and iteratively constructs the groundstate in terms of ``random singlets''. Although we do not expect this procedure to be accurate initially for finite disorder, we will see below that the effective disorder strength grows under renormalization, so that the method is said to be asymptotically exact -- {\it i.e.} is believed to give exact results for universal quantities. 

Let $\Omega = \max_i \lbrace J_i \rbrace$ be the largest coupling in the Hamiltonian, and let us parametrize the couplings as $\beta_i= \ln \frac{\Omega}{J_i}$. We also define a RG flow parameter $\Gamma = \ln\frac{\Omega_0}{\Omega}$ where $\Omega_0$ is the initial value of $\Omega$. Using the decimation rule~\eqref{eqdecimationrule} and ignoring a factor $\ln 2$ (irrelevant at strong disorder), one finds the flow equation for the distribution of the couplings $\beta_i$
\begin{align}
\frac{\partial P_{\Gamma}(\beta) }{\partial \Gamma}& = \frac{\partial P_{\Gamma}(\beta) }{\partial \beta} + P_\Gamma(0) P_\Gamma \star P_\Gamma (\beta), 
\end{align}
where $ P_\Gamma \star P_\Gamma (\beta) = \int d \beta_1   d \beta_2  P_\Gamma(\beta_1) P_\Gamma(\beta_2)  \delta \left( \beta -\beta_1 -\beta_2 \right)$ denotes the convolution. This equation has a remarkably simple solution~\cite{FisherRSRG2} that is essentially an attractor to all initial distributions 
\begin{equation}
P_\Gamma (\beta) = \frac{1}{\Gamma} {\rm e}^{-\beta/\Gamma}.
\end{equation}
Most features of random singlet critical points follow directly from this fixed point distribution, in particular the scaling~\eqref{eqScalingtE} between distance and energy, which can be recast as $\sqrt{\ell} \sim \Gamma$ where $\ell$ is the size of the singlets created at energy scale $\Gamma$. It is also possible to argue that even though the typical value of the spin-spin correlation function $\langle \vec{S}_0 . \vec{S}_r \rangle$ decays as ${\rm e}^{- c \sqrt{r}}$, its average decays much more slowly as $1/r^2$ as it is dominated by rare events where the two spins belong to the same singlet. It is also clear from this distribution that the effective disorder strength is $\Gamma$, and is therefore increasing at low energy.

\begin{figure*}
\subfloat[]{\label{figureCartoon0}  \includegraphics[width = 3in]{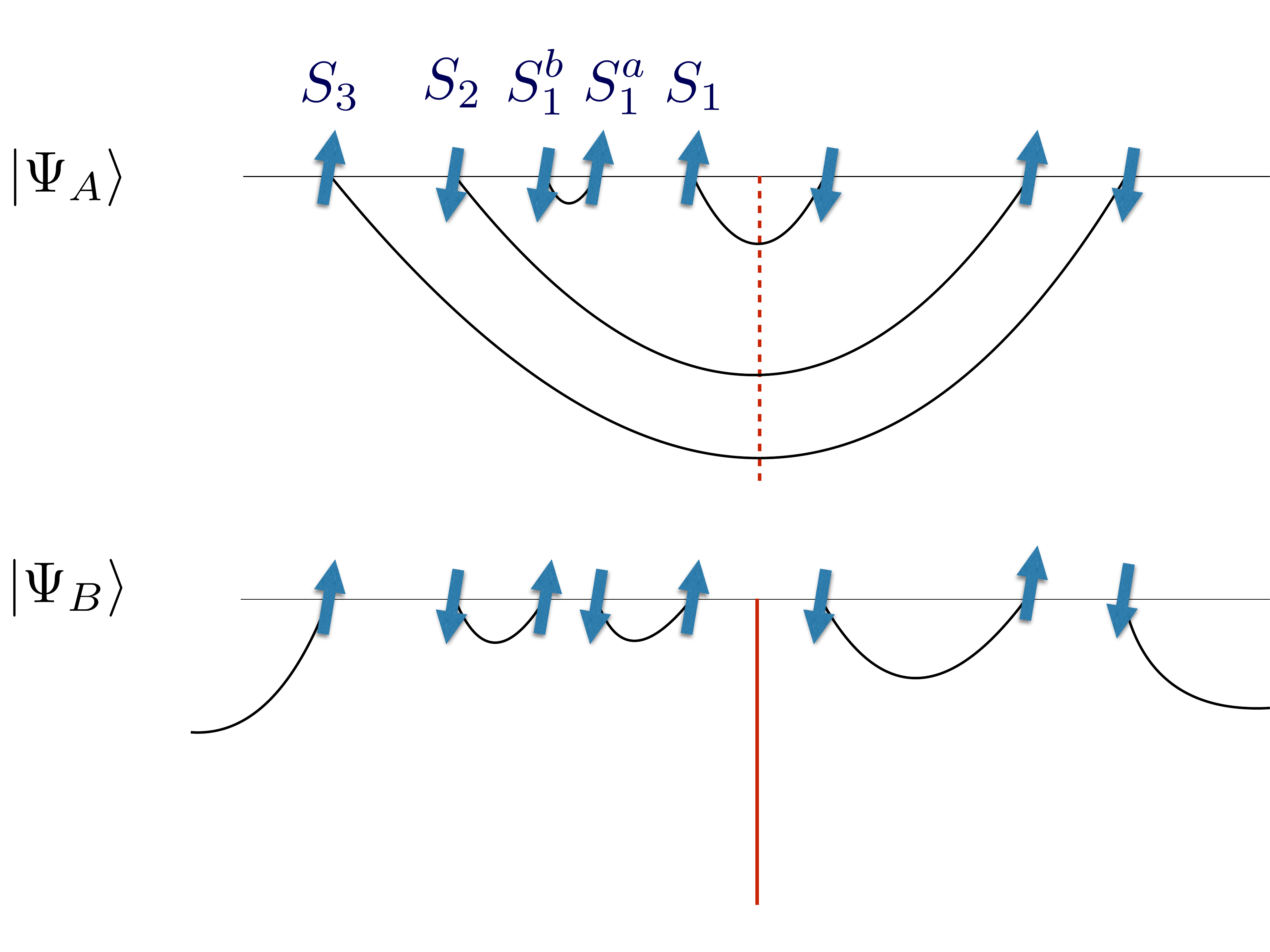}} \ \ \ \ \ \ \
\subfloat[]{\label{figureCartoon} \includegraphics[width = 3in]{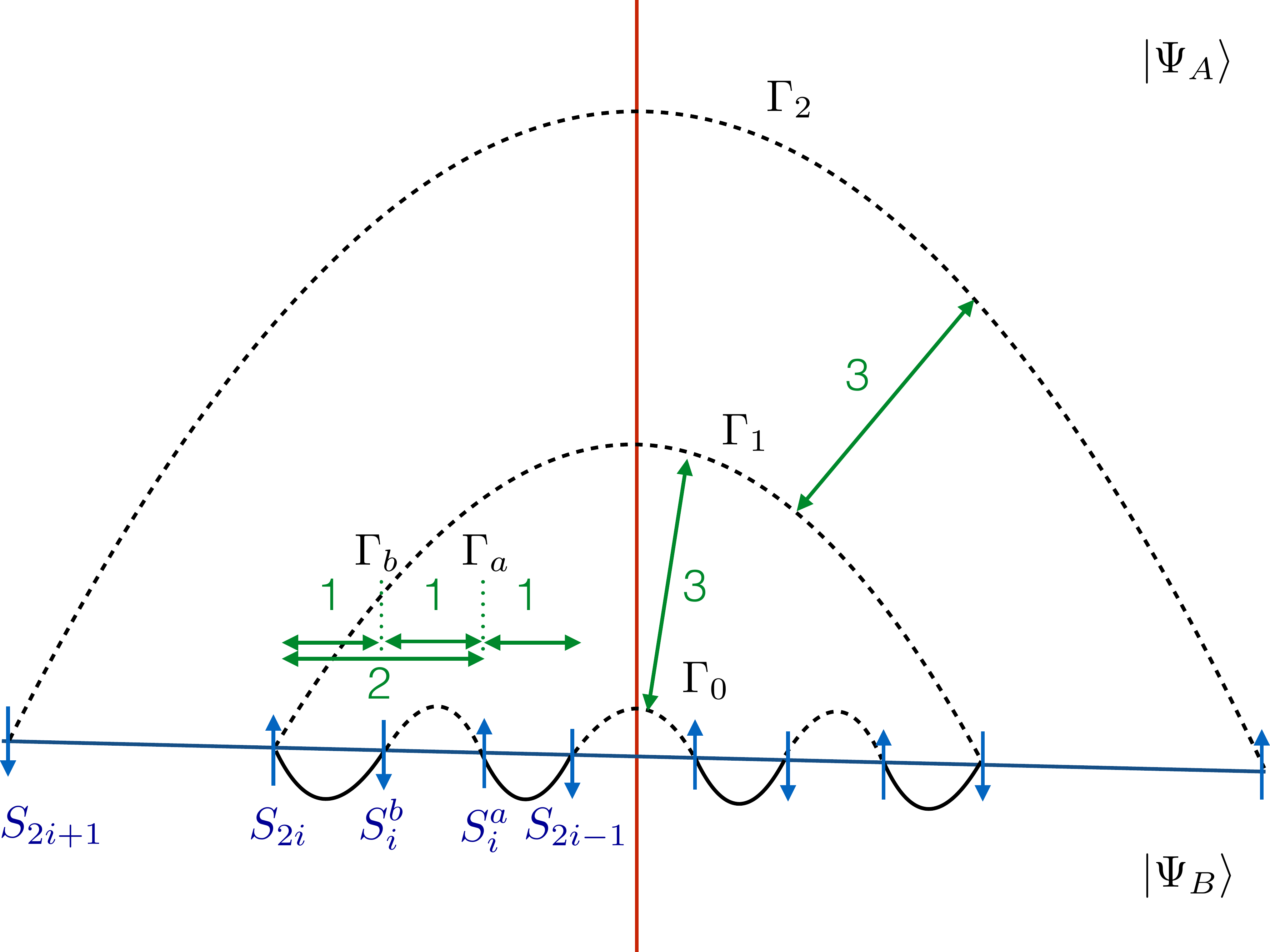}} 
\caption{(a) The singlets crossing the cut in $\Ket{\Psi_A}$ are obviously modified in $\Ket{\Psi_B}$, but other singlets can also be affected by the cut -- in this example, the singlet between the spins $S_1^a$ and $S_1^b$. (b) Schematic representation of the typical history dependence in the RG responsible for the scaling of $G_{\rm typ} \sim L^{-\ln 2/4}$. Dashed lines represent singlets of  $\Ket{\Psi_A}$ that are modified in  $\Ket{\Psi_B}$ (solid black lines) because of the cut. We also show in green the average RG times $\ln ( \Gamma_i/ \Gamma_j)$ between the creation of singlets at energy scales $\Gamma_i$ and $\Gamma_j$.}
\end{figure*}

\subsection{RSRG, entanglement entropy and groundstate overlap}

The random singlet structure of the groundstate can also be used to infer the scaling of more involved quantities of interest like entanglement entropy or overlaps. For example, considering a system of length $L=2N$ with $N$ even and open boundary conditions, the bipartite entanglement entropy between the right and left parts of the system (each of size $N$) is given by $S = n_e  \ln 2$ where $n_e$ is the number of singlets crossing the entanglement cut in the groundstate wavefunction $\Ket{\Psi_A}$. Using RSRG, the disorder average of this quantity was shown~\cite{RefaelMoore} to scale as $\overline{n}_e \sim \frac{1}{6} \ln L$ (here and in the following, $\overline{O}$ denotes the disorder average of $O$), leading to the critical-like scaling $\overline{S} \sim \frac{\ln 2}{6} \ln L$. The groundstate overlap $G =\Braket{\Psi_A | \Psi_B}$ can also be evaluated using the RSRG procedure: for a given distribution of disorder, the groundstates $\Ket{\Psi_A}$ and $\Ket{\Psi_B}$ without and with a cut in the middle of the system are given in terms of a collection of singlets (see left panel of Fig.~\ref{figureConfig}). The overlap then reads $G=\left(\frac{1}{2}\right)^{N_{\rm loops}-N/2}$, where $N_{\rm loops}$ is the number of loops in the configuration obtained by representing singlets by semicircles and by gluing the mirror image of $\Ket{\Psi_A}$ on top of $\Ket{\Psi_B}$ (see left panel of Fig.~\ref{figureConfig}). This quantity seems obviously more complicated than the number of singlets $n_e$ crossing the cut characterizing the entanglement entropy, but we will show that it is nevertheless possible to understand its scaling analytically in many cases. 

We remark that even though we consider open boundary conditions and a single cut in the middle of the system, the results below can be straightforwardly generalized to a system with periodic boundary conditions for $\Psi_A$, cut into two halves for $\Psi_B$: all the exponents derived below should then be multiplied by $2$ as this ``periodic" setup now involves two cuts instead of a single one.

\section{Typical decay of the groundstate overlap}
\label{sec2}

We first argue that the {\it typical} value of the overlap defined as $G_{\rm typ} \equiv {\rm e}^{\overline{\ln G}}$ decays with system size as a power law $G_{\rm typ} \sim L^{-\alpha_{\rm typ}}$  with $\alpha_{\rm typ} = \frac{\ln 2}{4}$.  Our starting point will be to express the overlap in terms of simpler observables whose statistics can be understood using the random singlet fixed point.  We first note that if the entanglement entropy of $\Ket{\Psi_a}$ is $0$ -- that is if the number of singlets crossing the boundary $n_e=0$ -- then $G=\Braket{\Psi_B |\Psi_A}=1$. In general however, $G$ is not simply related to $n_e$, although it is easy to show that $G \leq \left(\frac{1}{2} \right)^{n_e/2}$, where we recall that $n_e$ is even. This is consistent with the intuition that the wave-function overlap $G$ measures how different the wave-functions are with and without a cut, and has therefore no reason to be directly related to entanglement in general. For the random-singlet configurations that are generated by the RSRG, we find  that it is possible to express $G$ as
\begin{equation}
G = \left(\frac{1}{2}\right)^{n-n_e/2},
\label{eqGnne}
\end{equation}
where $n$ is the number of singlets by which $\Ket{\Psi_A}$ and $\Ket{\Psi_B}$ differ, and we recall that $n_e$ is the number of singlets crossing the cut in $\Ket{\Psi_A}$, so that the bipartite entanglement entropy in  $\Ket{\Psi_A}$ reads $S_A = n_e \ln 2$, while $S_B=0$ by definition. In other words, $n$ measures how many singlets are modified by the cut. Note that by definition, $n$ is larger than $n_e$ since singlets crossing the cut in  $\Ket{\Psi_A}$ have to be modified in $\Ket{\Psi_B}$, so that $n-n_e/2 \geq n_e/2$. We emphasize that eq.~\eqref{eqGnne} does not hold for arbitrary, generic random-singlet wave functions $\Ket{\Psi_A}$ and $\Ket{\Psi_B}$: it is an exact result that relies on the form of the (random-singlet) quantum states generated by the renormalization process and on the specific structure of the differences between $\Ket{\Psi_A}$ and $\Ket{\Psi_B}$ that we will describe below. 

We also note that intuitively, $G$ essentially behaves as a partition function or a correlation function -- this is actually how it can be calculated in the absence of disorder~\cite{PhysRevLett.101.120603, 1742-5468-2011-03-L03002,1742-5468-2011-08-P08019,PhysRevLett.110.240601} -- and contrary to say, the entanglement entropy, it is {\it not} self-averaging, with the natural quantity to average being $\ln G$ instead of $G$ itself. This intuition is confirmed by eq.~\eqref{eqGnne}, which shows that $G$ behaves in a way that is closely related to the {\it exponential} of the entanglement entropy, which we know should be self-averaging.

\subsection{Scaling of $\overline{n}_e$}

In order to understand the universal scaling of $G_{\rm typ}$, we need to compute the disorder average of $n$ and $n_e$ for a system of size $L$. The average $\overline{n}_e$ was computed in Ref.~\onlinecite{RefaelMoore}, where it was shown that the rate of singlet formations across the cut in $\Ket{\Psi_A}$ is given by
\begin{equation}
\label{eqRatene}
f(\mu) = \frac{1}{\sqrt{5}} \left({\rm e}^{-\frac{3-\sqrt{5}}{2} \mu}-{\rm e}^{-\frac{3+\sqrt{5}}{2} \mu} \right),
\end{equation}
where $\mu = \ln \frac{\Gamma}{\Gamma_0}$ is the total RG time between $\Gamma_0$ and  $\Gamma$. The average RG time between singlet formations is therefore given by $\overline{\mu} = \int_0^\infty d \mu f(\mu)  \mu = 3$, so that the number of singlets crossing the cut for a system of length $L$ reads $\overline{n}_e \sim \frac{\ln \Gamma}{3} \sim \frac{\ln L}{6}$ where we have used the random singlet scaling $\Gamma \sim \sqrt{L}$ between distance and energy. From this, we get the scaling of the entanglement entropy $S \sim \frac{\ln 2}{6} \ln L$ for a system with open boundary conditions~\cite{RefaelMoore}.

\subsection{History dependence of the RG}

The last ingredient we need is to understand how the average number of singlets modified by the cut $\overline{n} \geq \overline{n}_e$ scales with system size. In order to do this, we need to understand precisely the history dependence of the renormalization procedure, by following the RG flow in the configurations with (B) and without (A) the cut simultaneously.  The concept of ``history dependence'' of the RSRG was also used in Ref.~\onlinecite{RefaelMoore} and simply refers to the influence of a given decimation on future RG steps. For example, computing the factor $1/6$ in the scaling of the entanglement entropy $n_e \sim \frac{1}{6} \ln L$ requires taking into account the history of the singlet formations across the cut precisely: after being decimated, the renormalized central bond is typically much weaker (by a factor ${\rm e}^{-\Gamma}$) than the other bonds in the chain and thus has a much smaller probability of being decimated again. 

Let us imagine that $\Ket{\Psi_A}$ contains $n_e$ singlets crossing the cut, with $n_e$ even in our setup. For the same disorder configuration with the cut, let us focus on, say, the left half of the system and denote the spins that were involved in those $n_e$ singlets in $\Ket{\Psi_A}$ by $S_1,S_2, \dots, S_{2i-1},S_{2i},\dots, S_{n_e}$, where $i=1,\dots,n_e/2$ labels the spins by increasing distance to the cut. These spins will be reorganized in new singlets in $\Ket{\Psi_B}$, perhaps involving other spins that were not involved in singlets crossing the cut $\Ket{\Psi_A}$ (which is why $n$ can be larger than $n_e$). Focusing on the spins $S_{2i-1}$ and $S_{2i}$, two different scenarios can occur in $\Ket{\Psi_B}$: (i) $S_{2i-1}$ and $S_{2i}$ can form a singlet. (ii) $S_{2i-1}$ can form a singlet with another spin $S^{a}_{i}$ to its left, which was in turn involved in a singlet with another spin $S^{b}_{i}$ to its left in $\Ket{\Psi_A}$. This spin $S^{b}_{i}$ will have to form a singlet with a new spin to its left in $\Ket{\Psi_B}$, that can be either $S_{2i}$, or another intermediate spin $S^{c}_{i}$ that was in a singlet with a spin $S^{d}_{i}$ in $\Ket{\Psi_A}$ {\it etc} (see Fig.~\ref{figureCartoon0}). Importantly, the spins modified by the cut can occur only between the spins $S_{2i-1}$ and $S_{2i}$, not between $S_{2i}$ and $S_{2i+1}$ or $S_{2i-2}$ and $S_{2i-1}$. Note in particular that $S_{2i-2}$ and $S_{2i-1}$ or $S_{2i}$ and $S_{2i+1}$ cannot form singlets in $\Ket{\Psi_B}$, whereas $S_{2i}$ and $S_{2i-1}$ can. This particular structure is actually why~\eqref{eqGnne} holds.

\subsection{Scaling of $\overline{n}$}

We now estimate how many singlets in $\Ket{\Psi_A}$ $(S^{a}_{i}, S^{b}_{i}), (S^{c}_{i}, S^{d}_{i}) \dots$ between $S_{2i-1}$ and $S_{2i}$ are modified by the cut. We imagine running the RG until the scale $\Gamma_0$ at which $S_{2i-1}$ would form a singlet crossing the cut in $\Ket{\Psi_A}$. Denoting by $\Gamma_1 > \Gamma_0$ the scale at which $S_{2i}$ would form a singlet crossing the cut in $\Ket{\Psi_A}$, we know that the average RG time between those two events in $\Ket{\Psi_A}$ is given by $\overline{\ln \frac{\Gamma_1}{\Gamma_0}}=3$ (see above and Ref.~\onlinecite{RefaelMoore}). At the scale $\Gamma_0$ in the configuration B with the cut, the spin $S_{2i-1}$ is effectively at the right boundary of the left half of the system, and is coupled to its neighbor on the left by a bond of coupling strength $\beta$ given by the probability distribution $Q_{\Gamma_0}(\beta)=\frac{1}{\Gamma_0} {\rm e}^{-\beta/\Gamma_0}$. We now ask when this bond is decimated -- {\it i.e.} when $S_{2i-1}$ forms in singlet in $\Ket{\Psi_B}$. Following Ref.~\onlinecite{RefaelMoore}, we construct a flow equation for $Q_{\Gamma}(\beta)$ with the convention that $\int_0^\infty d \beta Q_{\Gamma}(\beta) = p_\Gamma$ be the probability that the bond involving $S_{2i-1}$ to its right was not yet decimated at scale $\Gamma$. We find that $Q_{\Gamma}(\beta)$  satisfies 
\begin{align}
\frac{\partial Q_{\Gamma}(\beta) }{\partial \Gamma}& = \frac{\partial Q_{\Gamma}(\beta) }{\partial \beta} + P_\Gamma(0) \left( P_\Gamma \star Q_\Gamma (\beta) - Q_{\Gamma}(\beta)  \right), 
\label{eqFlowEqn}
\end{align}
where the first term accounts for the change in $\beta$ when $\Gamma$ changes, and the second term corresponds to the flow of the coupling when the (single) neighbor of $S_{2i-1}$ forms a singlet. This equation is readily solved and we find
\begin{equation}
Q_\Gamma (\beta) = \frac{1}{\Gamma} {\rm e}^{-\beta/\Gamma} g \left(\mu = \ln \frac{\Gamma}{\Gamma_0} \right), \text{with } g(\mu)={\rm e}^{-\mu},
\label{eqQsinglesinglets}
\end{equation}
so that $p_{\Gamma} = g(\mu)$. The average RG duration after which the bond involving $S_{2i-1}$ is decimated in $\Ket{\Psi_B}$ is thus given by $\overline{\ln \frac{\Gamma_a}{\Gamma_0}} = \int_0^\infty d\mu g(\mu) \mu =1$ so that typically $\Gamma_1 > \Gamma_a > \Gamma_0$ where $\Gamma_a$ is the scale at which $S_{2i-1}$ forms a singlet with a spin $S^{a}_{i}$ in $\Ket{\Psi_B}$. We next consider this spin $S^{a}_{i}$ at scale $\Gamma_a$ in the configuration $A$ without the cut. $S^{a}_{i}$ is then coupled to its right neighbor by a coupling crossing the cut that will be decimated
\footnote{Note that this can be checked independently. The rate~\eqref{eqRatene} leading to the RG duration $\overline{\ln \frac{\Gamma_1}{\Gamma_0}}=3$ relies on two ingredients: (a) a flow equation for the central coupling similar to~\eqref{eqFlowEqn} with a factor 2 in front of the second term to account for the decimations of its two neighboring bonds, (b) the initial condition at scale $\Gamma_0$ at which the central bond was just decimated and is therefore strongly suppressed. At the scale $\Gamma_a > \Gamma_0$ however, it is natural to consider that the central bond is generic and is therefore distributed according to $Q_{\Gamma_a}(\beta)=\frac{1}{\Gamma_a} {\rm e}^{-\beta/\Gamma_a}$. Solving the flow equation starting from this initial condition at $\Gamma=\Gamma_a$, we find that $\overline{\ln \frac{\Gamma_1}{\Gamma_a}}=2$ consistent with  $\overline{\ln \frac{\Gamma_a}{\Gamma_0}}=1$.  } 
at scale $\Gamma_1$ with $\overline{\ln \frac{\Gamma_1}{\Gamma_a}}=2$, and to its left neighbor with a coupling distributed according to $Q_{\Gamma_a}(\beta)=\frac{1}{\Gamma_a} {\rm e}^{-\beta/\Gamma_a}$. Since $S^{a}_{i}$ is situated to the right of $S_{2i}$, the right bond cannot be decimated (otherwise it would form a singlet crossing the cut), and the average RG duration after which the bond left of $S^{a}_{i}$ is decimated in $\Ket{\Psi_A}$ is again given by $\overline{\ln \frac{\Gamma_b}{\Gamma_a}} = \int_0^\infty d\mu g(\mu) \mu =1$. This singlet in $\Ket{\Psi_A}$ involves $S^{a}_{i}$ and another spin $S^{b}_{i}$, that will in turn be at the right boundary of the left half of the system in the configuration B with the cut. By using the same argument, we find that $S^{b}_{i}$ will form a singlet $\Ket{\Psi_B}$ with a spin $S^{c}_{i}$ at a scale $\Gamma_c$ given by $\overline{\ln \frac{\Gamma_c}{\Gamma_b}}  =1$. However, because $\overline{\ln \frac{\Gamma_c}{\Gamma_0}} = \overline{\ln \frac{\Gamma_c}{\Gamma_b}} +\overline{\ln \frac{\Gamma_b}{\Gamma_a}} +\overline{\ln \frac{\Gamma_a}{\Gamma_0}}  =1+1+1=3$, we can identify $\Gamma_c=\Gamma_1$ and $S^{c}_{i} = S_{2i}$ so that the cycle stops. After the scale $\Gamma_1$ at which $S_{2i}$ forms a singlet crossing the cut in $\Ket{\Psi_A}$,  all the decimations are identical in the configurations A and B until the scale $\Gamma_2$ (given typically by $\overline{\ln \frac{\Gamma_2}{\Gamma_1}}=3$) at which another singlet crosses the cut in $\Ket{\Psi_A}$.

We therefore end up with a picture of the typical history dependence summarized in Fig.~\ref{figureCartoon}, where within a total RG time $\overline{\ln \frac{\Gamma_2}{\Gamma_0}}=6$, $n=4$ singlets were modified by the cut: $n_e=2$ singlets crossing the cut in $\Ket{\Psi_A}$, as well as two additional singlets involving the spins $S^{a}_{i}$ and $S^{b}_{i}$ on the left half of the system, and similarly on the right half. Therefore, we find $\overline{n} \sim 4 \frac{\ln \Gamma}{6} \sim \frac{1}{3} \ln L \sim 2 \overline{n}_e$, up to non-universal contributions. 

\subsection{Scaling of $G_{\rm typ}$}
\label{subsecTyp}

Gathering these different ingredients, we are ready to compute the average $\overline{\ln G} = -\ln 2 \left( \overline{n} -  \overline{n}_e/2 \right)$. Our results then imply a power-law decay of the typical value of $G$ given by
\begin{equation}
G_{\rm typ} \sim L^{-\ln 2 / 4}.
\label{eqGtypmain}
\end{equation}

Note that although we have focused our analysis on the Heisenberg~\eqref{eqXXX} and XXZ chains~\eqref{eqXXZ}, the results can be readily generalized to other random singlet spin chains. For instance, let us consider the random bond transverse field Ising chain
\begin{equation}
H= - \sum_i J_i \sigma_i^z \sigma_{i+1}^z + h_i \sigma_i^x, 
\label{eqIsing}
\end{equation}
where $J_i$ and $h_i$ are random with $\overline{\ln h_i}=\overline{\ln J_i}$, corresponding to a random quantum critical point separating a paramagnetic phase and a ferromagnetic phase. This critical point can be conveniently interpreted as a random singlet critical point where the singlets are formed between Majorana fermions~\cite{PhysRevLett.99.140405}. The only difference in our analysis is that~\eqref{eqGnne} now reads $G = \left(\frac{1}{\sqrt{2}}\right)^{n-n_e/2}$, where $d=\sqrt{2}$ can be interpreted as the {\it quantum dimension} of a Majorana fermion, whereas we had $d=2$ for the spin-$\frac{1}{2}$ chains~\eqref{eqXXX} and~\eqref{eqXXZ}. Therefore, the typical value of the overlap $G$ for the critical random Ising chain~\eqref{eqIsing} scales as
\begin{equation}
G^{\rm Ising}_{\rm typ} \sim L^{-\ln 2 / 8}.
\end{equation}
More generally, our result applies directly to the random $SU_k(2)$ anyonic chains studied in Ref.~\onlinecite{PhysRevLett.99.140405} by replacing factors of $2$ (for Heisenberg) or $\sqrt{2}$ (for Ising) by the appropriate quantum dimension. 

It is also instructive to compare our result to the pure (disorder free) case, where the overlap between the wave functions with and without a cut can be computed using Conformal Field Theory (CFT)~\cite{1742-5468-2011-03-L03002}. The overlap then goes to zero in the thermodynamic limit as $G\sim L^{-c/16}$, where $c$ is the {\it central charge} of the corresponding CFT -- with $c=1$ for the Heisenberg chain or $c=1/2$ for the Ising chain. Note in particular that our results for the random case are {\it not} obtained directly from the pure case by replacing $c$ by the ``effective central charge'' introduced in Ref~\onlinecite{RefaelMoore} for the entanglement entropy.

\section{Multifractal Orthogonality Catastrophe Spectrum}
\label{sec3}

In the previous section, we showed that the typical value $G_{\rm typ}$ of the wavefunction overlap $G=\Braket{\Psi_A | \Psi_B}$ suffers from an Anderson orthogonality catastrophe and decays as a powerlaw of the system size. It is natural to ask whether this orthogonality catastrophe also holds for the mean value $\overline{G}$. More generally, it is interesting to study the behavior of $\overline{G^\alpha}$ with $\alpha>0$. In this section, we will show that 
\begin{equation}
\overline{G^\alpha} \sim L^{- \mu(\alpha)},
\end{equation}
where the multifractal exponent $\mu(\alpha)$ depends very non-trivially on $\alpha$, with in particular an interesting limit $\mu(\infty)$. We emphasize that we are using the word ``multifractal'' in a broad sense here, meaning that the exponent $\mu(\alpha)$  depends non-linearly on $\alpha$. 

\subsection{$\alpha \to 0$ and $\alpha \to \infty $ limits}
 
Using eq.~\eqref{eqGnne}, we first express $\overline{G^\alpha}$ as 
\begin{equation}
\overline{G^\alpha} = \sum_{n,n_e} P(n,n_e) \left(\frac{1}{2}\right)^{\alpha (n-n_e/2)},
\label{eqMultiFract}
\end{equation}
where $P(n,n_e)$ is the probability to have a random singlet configuration with $n $ singlets modified by the cut, and $n_e$  singlets crossing the cut in $\Ket{\Psi_A}$. There are two limits that can be understood easily. First of all, in the limit $\alpha \to 0$, $\overline{G^\alpha} \simeq G_{\rm typ}^\alpha \sim L^{-\alpha \ln 2 / 4} $, so that 
\begin{equation}
\mu(\alpha) \underset{\alpha \to 0}{\sim} \alpha \frac{\ln 2}{4},
\label{eqLimit1}
\end{equation}
using the results of the previous section. 

It is also interesting to study the opposite limit $\alpha \to \infty$. As $\alpha \to \infty$,  the only configurations that survive in eq.~\eqref{eqMultiFract} are those with $n=0$ and $n_e=0$. That is to say, the limit $\alpha \to \infty$ counts the number of random singlet configurations that are completely unaffected by the cut: $G=1$ and $S_A=0$. The number of such configurations scales with a non-trivial exponent that can be deduced~\cite{RefaelMoore, PhysRevB.83.045110} from~\eqref{eqRatene} for example: $P(n=0,n_e=0) \sim L^{-(3-\sqrt{5})/4}$. This critical exponent can be rewritten as $1-\frac{\varphi}{2}$ where $\varphi = \frac{1+\sqrt{5}}{2}$ is the golden ratio. The golden ratio appears in many different quantities in random singlet critical points, all of which being related in one way or another to the probability that a given bond be not decimated in the course of the RG. This yields
\begin{equation}
\mu(\alpha) \underset{\alpha \to \infty}{\to} \frac{3 -\sqrt{5}}{4}.
\label{eqLimit2}
\end{equation}
For $\alpha \gg 1$, we therefore find $\overline{G^\alpha} \sim L^{-\frac{3 -\sqrt{5}}{4}}$, which is dramatically different from the very small typical scaling $G^\alpha_{\rm typ}  \sim L^{-\alpha \ln 2 / 4}$. In this limit, the average $\overline{G^\alpha} $ is thus dominated by rare configurations with $G=1$ that are unaffected by the cut.

Of course, in general $\overline{G^\alpha} \sim A(\alpha) L^{- \mu(\alpha)}$ where $ A(\alpha) $ is a non-universal constant that will vanish in the limit $\alpha \to \infty$ so that strictly speaking, $\lim_{\alpha \to \infty } \overline{G^\alpha} = 0$. However, as we will see below, $\overline{G^\alpha} $ is dominated by rare configurations (and thus decays with an exponent $\approx \frac{3 -\sqrt{5}}{4}$) already for $\alpha \gtrsim 10$.

\subsection{Multifractal Anderson Orthogonality exponents and generating functions}

We now turn to the calculation of the full multifractal spectrum $\mu(\alpha)$. In order to do so, we need to compute the full generating function ${\cal F}(u,v)=\sum_{n,n_e} P(n,n_e) u^n v^{n_e}$ of the joint distribution $P(n,n_e)$, with $\overline{G^\alpha} = {\cal F}\left( \frac{1}{2^\alpha}, \frac{1}{2^{-\alpha/2}} \right)$. This is obviously a very hard problem, and we expect the calculation below to be only approximate -- whereas our results up to this point should be exact. 

The generating function ${\cal G}(v)={\cal F}(1,v)=\sum_{n_e} P_{n_e} v^{n_e}$ is related to the entanglement spectrum and was computed (in an approximate way) in Ref.~\onlinecite{PhysRevB.83.045110}. The key idea was to assume that this generating function satisfies the following renewal equation as a function of RG time $\mu$
\begin{equation}
{\cal G}_\mu(v)= \int_{\mu}^\infty d \mu^\prime f(\mu^\prime) + v \int_0^{\mu} d \mu^\prime f(\mu^\prime) {\cal G}_{\mu-\mu^\prime}(v),
\label{eqRenewalEq}
\end{equation}
where the singlet formation rate $f(\mu)$ is given by~\eqref{eqRatene}. This equation can be readily solved by Laplace transform, and one finds
\begin{equation}
{\cal G}(v) \sim L^{-\left(\frac{3-\sqrt{5+4 v}}{4} \right)},
\label{eqgenfne}
\end{equation}
where we have used the fact that $\ln \Gamma \sim \frac{1}{2} \ln L$. In particular, one can check that $\overline{n}_e = {\cal G}^\prime(v=1)=\frac{1}{6} \ln L$. We emphasize that the renewal equation~\eqref{eqRenewalEq} is ignoring memory effects beyond first order, multiple decimations {\it etc}. It is actually possible to show numerically that~\eqref{eqgenfne} does not hold exactly (see below), even if it provides a very good approximation of the exact result.   

The calculation of the generating function ${\cal F}(u,v)$ is even more intricate. To do this, we assume that the joint probability distribution $P(n,n_e)$ factorizes as $P(n,n_e) \approx P(n_e) H(n-n_e)$, where $H(n-n_e)$ is the probability distribution for the number of singlets modified by the cut that were not crossing the cut in $\Ket{\Psi_A}$. Those events are governed by the rate function $g(\mu)= {\rm e}^{- \mu}$ in eq.~\eqref{eqQsinglesinglets}, and should be described by a Poisson process since they are essentially independent. For a configuration with $n_e$ singlets crossing the cut, there are $n_e/2$ ``active regions'' where $n-n_e$ can increase, and we have seen above that typically, each one of these regions has $n=4$ singlets modified by the cut, so that $n-n_e=2$ for each active regions. We thus take $H(n-n_e)$ to be a Poisson process with parameter $n_e/2 \times 2 = n_e$, so that $P(n,n_e) \approx P(n_e) \frac{{\rm e}^{-n_e}}{(n-n_e) !} n_e^{n-n_e}$. We can now evaluate the generating function ${\cal F}(u,v)$ as
\begin{equation}
{\cal F}(u,v) \approx \sum_{n_e} P(n_e) \left( u v {\rm e}^{u-1}\right)^{n_e} = {\cal G}( u v {\rm e}^{u-1}),
\end{equation}
so that the full generating function decreases as a power law as
\begin{equation}
{\cal F}(u,v) \approx L^{-\left(\frac{3-\sqrt{5+4 u v {\rm e}^{(u-1)}}}{4} \right)}.
\end{equation}
One can check that this formula gives $\left.\partial_u {\cal F}(u,v) \right|_{u,v=1} = \overline{n}=\frac{1}{3} \ln L  $ and $\left.\partial_v {\cal F}(u,v) \right|_{u,v=1} = \overline{n}_e=\frac{1}{6} \ln L  $ as it should. Going back to the multifractal Anderson orthogonality spectrum, this gives
\begin{equation}
\mu(\alpha) \approx \frac{3-\sqrt{5+4 \frac{1}{(\sqrt{2})^\alpha} \exp{\left(2^{-\alpha}-1\right)}}}{4}.
\label{eqGmultspect}
\end{equation}
This formula is compatible with the exact limits~\eqref{eqLimit1} and~\eqref{eqLimit2}.

\section{Numerical Results}
\label{sec4}


\begin{figure*}
\subfloat[]{\label{figureNum}  \includegraphics[width = 3.3in]{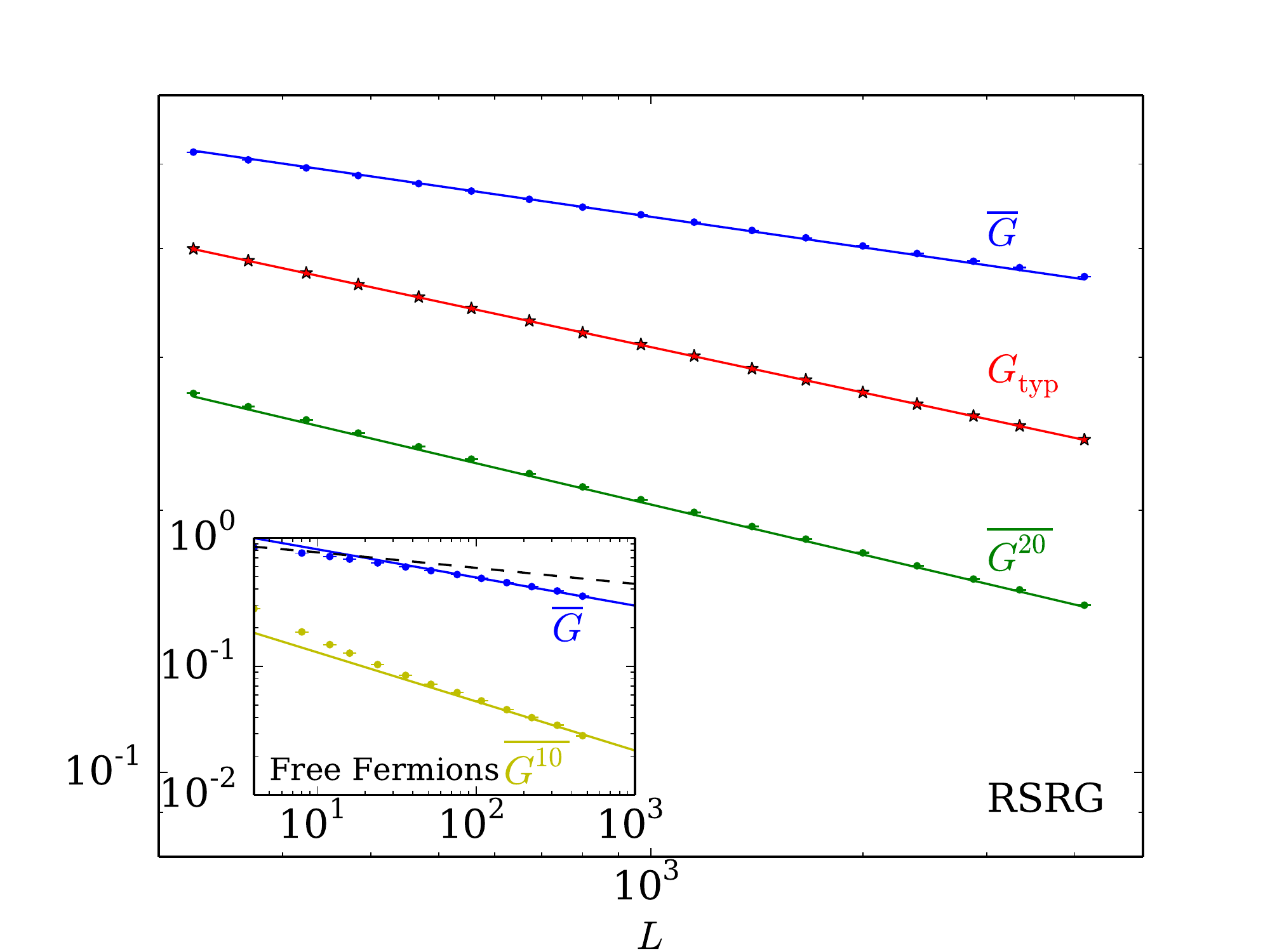}} \ \ \ \ \ \ \
\subfloat[]{\label{figureFractalAOC} \includegraphics[width = 3.3in]{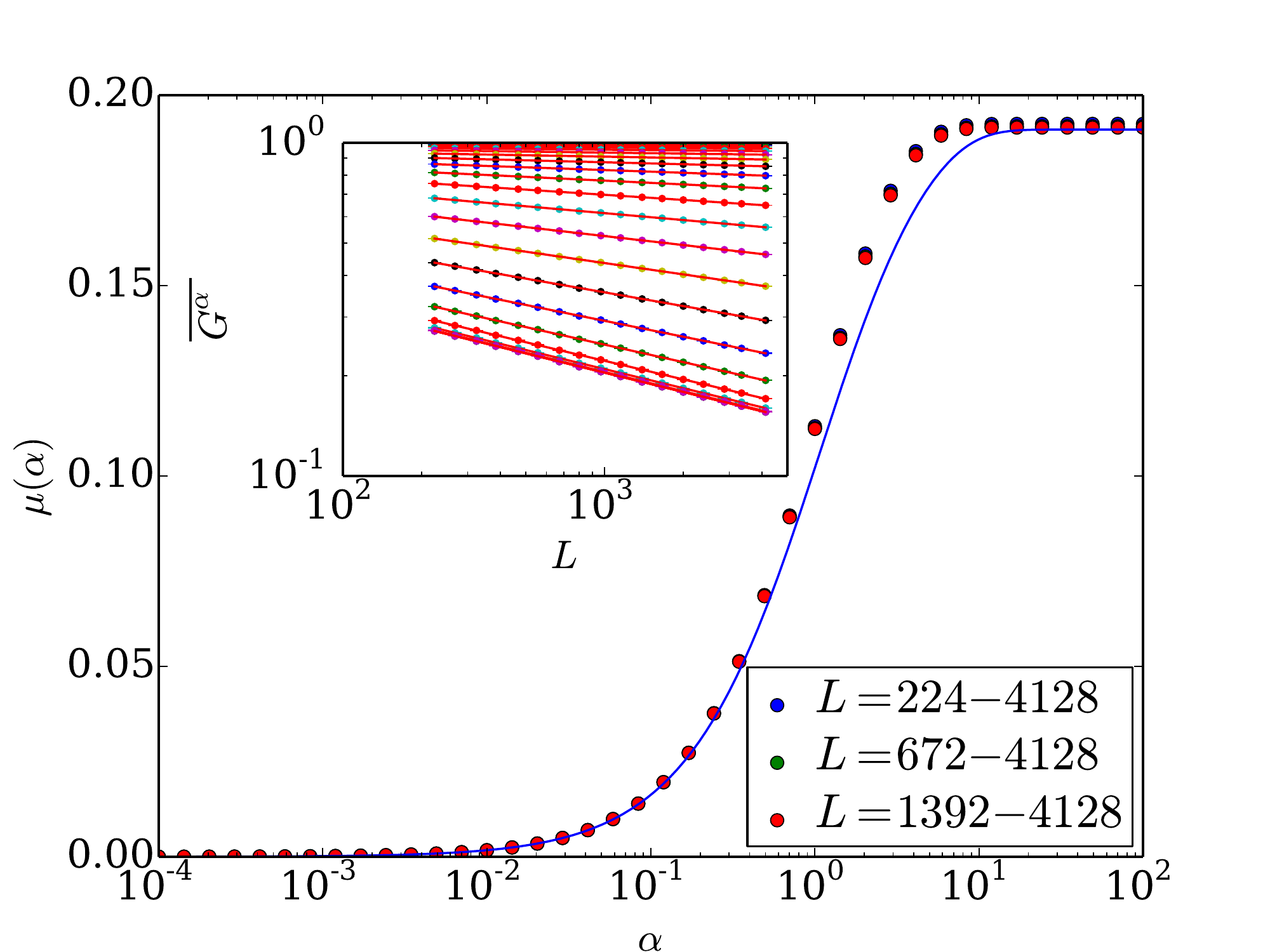}} 
\caption{(a) Power law dependence of $G_{\rm typ}$ and $\overline{G^\alpha}$ as a function of system size $L$. The data points  obtained from a numerical implementation of the renormalization group agree very well with our analytic predictions (solid lines). Inset: Numerical check of this power-law dependence for a random-bond XX chain using free fermion techniques (solid lines: RSRG predictions, dashed line: overlap for a clean chain). (b) Multifractal Orthogonality spectrum $\overline{G^\alpha} \sim L^{- \mu(\alpha)}$. The numerically extracted exponent $\mu(\alpha)$ agrees quite well with eq.~\eqref{eqGmultspect} (blue line). Inset: $\overline{G^\alpha}$ against $L$ for various values of $\alpha=10^{-4}, \dots, 10^2$. }
\end{figure*}

In this section, we check our results numerically confirming that random singlet critical points suffer from an Anderson orthogonality catastrophe. Admitting that the RSRG provides an accurate description of the groundstate of the random chains~\eqref{eqXXX} and~\eqref{eqXXZ}, the universal scaling of the overlap $G$ can be very efficiently computed by implementing the RSRG procedure numerically. Starting from a given disorder configuration, we identify and decimate the strongest bonds by forming singlets. We repeat this procedure until we spanned the whole chain, and we end up with the groundstate wave function consisting of a collection of singlets. We compute the groundstates with and without the cut for each disorder realization,  and we compute the overlap $G$ as explained in Sec.~\ref{sec1}. We draw the random couplings $J_i \in [0,1]$ from the distribution $P(J)=\frac{1}{W} \frac{1}{J^{1-1/W}} $ with $W=5$. Results for different system sizes $L$  are shown in Fig.~\ref{figureNum}, averaged over more than $2 \times 10^6$ disorder realizations. We find that $G_{\rm typ}$, $\overline{G}$ and $\overline{G^{20}}$ clearly vanish algebraically with $L$, and the corresponding critical exponents are found to be in very good agreement with~\eqref{eqGtypmain} and~\eqref{eqGmultspect}. In particular, $\overline{G^{20}}$ decreases with $L$ very slowly with an exponent close to~\eqref{eqLimit2}. By fitting $\overline{G^{\alpha}}$ as a function of $L$, we also measured the exponent $\mu(\alpha)$ for various values of $\alpha$ (Fig.~\ref{figureFractalAOC}). The numerical results are in relatively good agreement with the (approximate) formula~\eqref{eqGmultspect}, but although the finite size corrections seem to tend in the right direction, eq.~\eqref{eqGmultspect} is clearly slightly off for $\alpha \sim 1$. We find similarly that eq.~\eqref{eqgenfne} for the entanglement spectrum~\cite{PhysRevB.83.045110} is ruled out by numerical results, despite being very close to the exact solution. We leave the possibility of deriving exact formula for these quantities for future work.  

Our results can also be checked numerically for the random XX chain ($\Delta=0$ in eq.~\eqref{eqXXZ}) that can be mapped onto free fermions after a Jordan-Wigner transformation. Since the Jordan-Wigner transformation is nonlocal, in principle one has to be careful when comparing real space properties. However, the key quantities we study, such as entanglement and overlaps, are independent of whether one chooses the spin or fermionic representation; at most there are constant factors related to the boundary condition used in defining the Jordan-Wigner string, but these do not modify the scaling behavior that is the focus of our study.  We point out that our results apply equally well to the fermionic problem. For a free fermion Hamiltonian $H=\sum_{ij} h_{ij} c^\dagger_i c_j$, the groundstate wavefunction can be obtained by diagonalizing the matrix $h_{ij}$ and by filling the Fermi sea. The overlap $G$ between two different groundstates can then be expressed as a determinant using Wick's theorem. Unfortunately, we find that the numerical evaluation of such determinants becomes unstable at strong disorder and/or large distances -- that is, in the random singlet regime we are interested in. We note that even though similar free fermions methods were used to check the scaling of the entanglement entropy in the random singlet regime~\cite{PhysRevB.72.140408}, the computation of the entanglement entropy is also plagued by numerical instabilities at strong disorder\footnote{N. Laflorencie: private communication.}. Groundstate overlaps seem unfortunately even more sensitive to such instabilities, making it hard to access the random singlet physics. Nevertheless, by restricting our numerics to a regime (corresponding to moderate system size $L$) where the instabilities are negligible, we find a good agreement between numerical results for couplings drawn from a uniform distribution $J_i \in [0,1]$ and our analytic predictions (see inset in Fig.~\ref{figureNum}).


\section{Generic perturbations and local quantum quenches}
\label{sec5}

In this section, we go back to more general arbitrary local perturbations and conjecture the form of the dynamical response of a random singlet critical point to a sudden local perturbation (local quantum quench). 

\label{secQuenches}

We mostly focused above on a specific type of local perturbation that was amenable to analytic calculations: namely, a cut in the system. It is also instructive to consider more generic local perturbations. For example, we also computed (by implementing the RSRG procedure numerically) the overlap $G=\Braket{\Psi_A | \Psi_B}$ between the groundstate  $\Psi_A$ of a disordered spin chain of size $L=2N$ with random couplings $J_1, \dots, J_N, \dots J_{L-1}$, and the groundstate  $\Psi_B$ of the same disorder configuration with a weakened central bond $\tilde{J}_N = \lambda J_N$ with $\lambda<1$ (the ``cut'' case considered in most of this paper then corresponding to $\lambda=0$). We find that the system suffers from an Anderson orthogonality catastrophe only in the ``cut'' case $\lambda=0$. In other words, the nature of this algebraic decay of the groundstate overlap seems to be very different from the clean case where any arbitrary weak local perturbation is enough to induce an  Anderson orthogonality catastrophe. For random singlet systems, we find that the overlap $G$ remains finite in the thermodynamic limit $L \to \infty$ for any value of $\lambda \neq 0$.

   \begin{figure}[t!]
\includegraphics[width=1.0\columnwidth]{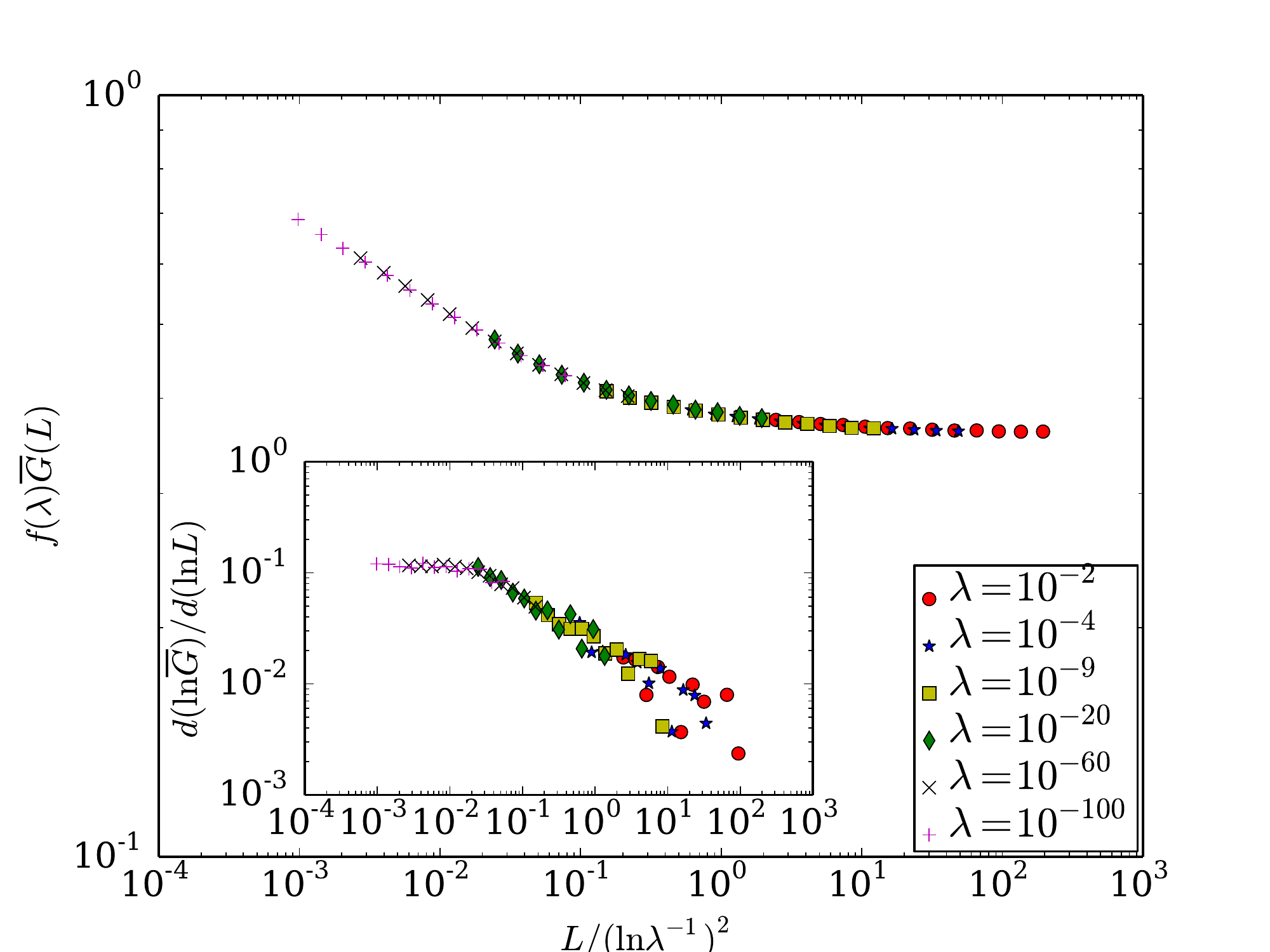}
\caption{Crossover physics of the overlap $\overline{G}(L)$ for a local perturbation corresponding to a weakened link $\tilde{J}_N = \lambda J_N$ with $\lambda<1$. Once plotted against  $L/L^\star \sim L/ \left(\ln \lambda^{-1}\right)^2$, we find that the data for the overlap collapse up to a rescaling by a non-universal prefactor $f(\lambda)$ that we adjust freely. Inset: Universal collapse of the logarithmic derivative $ \frac{d \ln \overline{G} }{d \ln L}$ (no free parameter). }
\label{figureCrossover}
\end{figure}

\subsection{Crossover for generic local perturbations}
\label{subsecLocalPe}

Using the scaling of the random singlet fixed point, we introduce a characteristic length scale $L_\lambda \sim \left(\ln \lambda^{-1}\right)^{1/\psi}$ with $\psi=\frac{1}{2}$ associated with the energy scale of the weakened link.  At high energies $\Omega \gg \tilde{J}_N$ (corresponding to system sizes $L \ll L_\lambda$), the weakened bond can be essentially considered as $\tilde{J}_N \approx 0$, and our results for the cut should apply in this regime with the overlap $\overline{G^\alpha}$ scaling as $L^{-\mu(\alpha)}$. This is very intuitive physically: if $L$ is small enough (or equivalently if $\lambda$ is small enough) so that $L \ll L_\lambda$, the central bond with strength $\tilde{J}_N$ will almost surely be the last bond decimated during the RSRG procedure so there is no difference between a weakened bond and an actual cut. On the other hand, as the cutoff $\Omega$ is lowered and becomes of order $\tilde{J}_N$ ($L \sim L_\lambda$), the fact that the weakened bond is in fact non-zero becomes important, and the overlap $\overline{G}$  goes to a (non-universal) constant that depends on $\lambda$ as $L \gg L_\lambda$. We verify this crossover scaling by collapsing $\overline{G}(L)$ curves (up to a global non-universal prefactor $f(\lambda)$ that we adjust freely) for different values of $\lambda$ (see Fig.~\ref{figureCrossover}). To get rid of this non-universal $\lambda$-dependent part in $\overline{G}(L)$, we also checked that the logarithmic derivative $ \frac{d \ln \overline{G} }{d \ln L}$ collapses onto a single universal curve without any free parameter once plotted against $L/L^\star \sim L/ \left(\ln \lambda^{-1}\right)^2$ (see inset in Fig.~\ref{figureCrossover}).

\subsection{Local quantum quenches}
 
 Let us also briefly comment on the generalization of our results to out-of-equilibrium setups. Let $\Ket{\Psi^{(n)}_A}$ be the (sorted) eigenstates of the Hamiltonian $H_A$ of a random spin chain and let $\Ket{\Psi^{(n)}_B}$ be the eigenstates of the Hamiltonian $H_B=H_A + V$ of the same system with a local perturbation $V$. Starting from the initial state $\Ket{\Psi^{(0)}_A}$, we imagine a quench protocol where the local perturbation is suddenly turned on at time $t=0$, so the wave-function of the system at time $t$ reads $\Ket{\Psi(t)}={\rm e}^{-i H_B t} \Ket{\Psi^{(0)}_A} $. The time-dependent analog of the groundstate overlap $G=\Braket{\Psi^{(0)}_A \left| \Psi^{(0)}_B} \right.$ is then known as the Loschmidt echo
\begin{equation}
G(t)=\Braket{\Psi^{(0)}_A \left| {\rm e}^{i H_A t} {\rm e}^{-i H_B t}\right|\Psi^{(0)}_A}, 
\label{eqLosch}
\end{equation}
which up to a phase is nothing but the overlap of the wavefunction at time $t$ with the initial wavefunction at time $t=0$. We will consider the case where the local quench corresponds to either two random spin chains suddenly glued together, or to a single random spin chain suddenly cut in half at time $t=0$.

For pure gapless spin chains, the Loschmidt echo after such a local quench decays~\cite{1742-5468-2011-08-P08019} as a power-law $G(t) \sim t^{-\eta}$ where $\eta$ is the same exponent as the orthogonality exponent characterizing the decay of the ``static'' groundstate overlap $G \sim L^{-\eta}$ as a function of the system size~\cite{1742-5468-2011-03-L03002} -- this can essentially be traced back to the symmetry between time and space (dynamical exponent $z=1$) of these systems. The Loschmidt echo seems quite hard to compute using RSRG but it is natural to expect the behavior of $G(t)$ in the time dependent setup to be related to the static scaling $\overline{G^\alpha} \sim L^{-\mu(\alpha)}$ using the low-energy relation $\ln t \sim \sqrt{L}$ (see eq.~\eqref{eqScalingtE} with $t^{-1}=E$ acting as an energy scale). We therefore conjecture that for local quantum quenches, the long-time behavior of the Loschmidt echo should be given by
\begin{equation}
\overline{G(t)^\alpha} \sim \frac{1}{ \left( \ln t\right)^{2 \mu(\alpha)}},
\label{eqGt}
\end{equation}
where $\mu(\alpha)$ is the same exponent as the one appearing in the static groundstate overlap $\overline{G^\alpha} \sim L^{-\mu(\alpha)}$. Using the same argument for the entanglement entropy yields a very slow growth $S(t) \sim \ln \ln t$ after a local quench, consistent with Ref.~\onlinecite{PhysRevB.85.094417}. Note that the behavior after a global quench would be dramatically different: considering the entanglement entropy for instance, the entanglement growth starting from a product state for the random Ising chain becomes $S(t) \sim (\ln t)^\alpha$ in the presence of interactions breaking integrability~\cite{PhysRevLett.112.217204}.

\begin{figure}[t!]
\includegraphics[width=1.0\columnwidth]{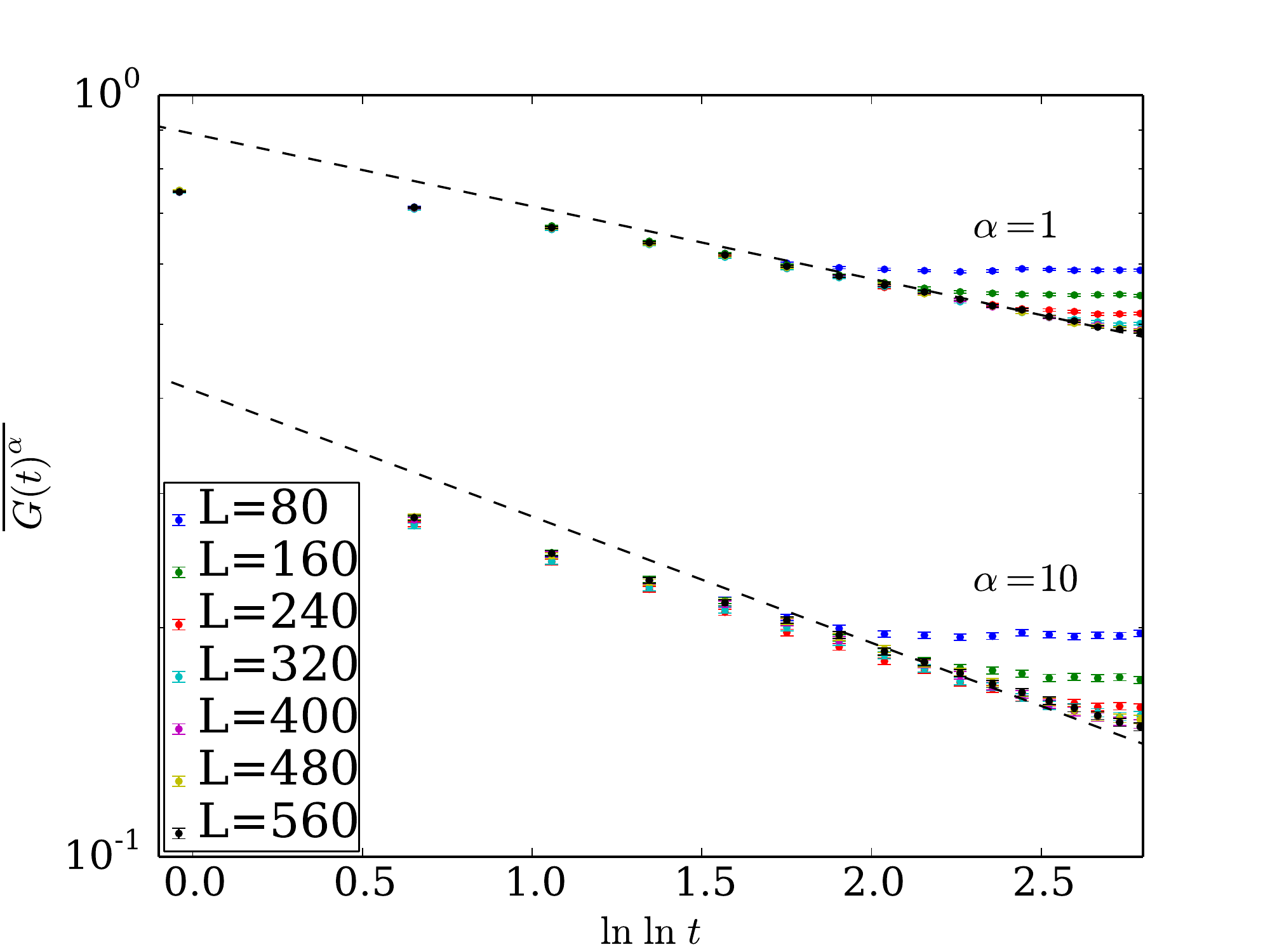}
\caption{Slow decay of the Loschmidt echo after two random XX chains of size $L/2$ are suddenly connected by a random bond (local quench). }
\label{figureQuench}
\end{figure}

 We checked the scaling~\eqref{eqGt} numerically in the random XX chain (with couplings drawn from the uniform distribution on $[0,1]$) where the local quench consists in suddenly connecting two chains of size $N$ by a random coupling drawn from the same distribution (Fig.~\ref{figureQuench}). By computing the Loschmidt echo as a determinant using Wick's theorem, we found a long time behavior of $\overline{G(t)^\alpha} $ compatible with eq.~\eqref{eqGt}, with the exponent $\mu(\alpha)$ consistent with our results in the static case. 

\label{secQuenches}

\subsection{Work distribution}

Another quantity of interest is the Fourier transform ${\cal A} (\omega) $ of the Loschmidt echo~\eqref{eqLosch} $G(t)=\int d\omega {\rm e}^{-i \omega t}{\cal A} (\omega)  $ that corresponds physically to the distribution of the work done during the quantum quench~\cite{PhysRevLett.101.120603,PhysRevLett.109.250602} 
\begin{equation}
{\cal A} (\omega) = \sum_n \left| \Braket{\Psi^{(n)}_B |\Psi^{(0)}_A} \right|^2 \delta\left(\omega - E^{(n)}_1+E^{(0)}_0 \right).
\end{equation}
In optical absorption experiments realizing quantum quenches, ${\cal A} (\omega) $ corresponds to the absorption spectrum where the energy $\hbar \omega$ of a photon naturally coincides with the work needed to perform the quantum quench~\cite{PhysRevLett.106.107402,Latta:2011aa,PhysRevLett.108.190601,PhysRevLett.110.230601}. In clean systems, the power-law decay of the Loschmidt echo with time $G(t) \sim t^{-\eta}$ then translates into an edge singularity ${\cal A} (\omega)  \sim \theta(\omega) \omega^{\eta-1}$ similar to the X-ray singularity in metals~\cite{PhysRev.163.612,PhysRev.178.1097}, with $\theta(x)$ the Heaviside step function. Notice that in this expression, we set  the minimal work required to perform the quantum quench $\omega_0=E^{(0)}_1-E^{(0)}_0$ to zero by convention -- this can always be done by shifting the frequencies $\omega-\omega_0 \to \omega$. For our local quench consisting of two random singlet spins chains glued together at time $t=0$, the average Loschmidt echo is given by eq.~\eqref{eqGt} with $\alpha=1$ so that the work distribution (absorption spectrum) scales at low frequencies as
\begin{equation}
\overline{{\cal A}} (\omega) \sim \frac{1}{\omega} \frac{1}{\left( \ln \frac{1}{\omega} \right)^{1+2 \eta}},
\end{equation}
with $\eta=\mu(1)$. Note that this singularity is integrable, compatible with the fact that $\int d \omega \overline{{\cal A}} (\omega)  =1$ is finite. The very slow decay~\eqref{eqGt} of the Loschmidt echo thus translates into a highly singular (barely integrable) behavior at low frequencies. It would be interesting if the methods of Refs.~\onlinecite{PhysRevLett.84.3434, PhysRevB.63.134424} to compute equilibrium dynamical response functions in random spin chains could be extended to recover this result.  

~
 
\section{Conclusion}
\label{sec6}

We have studied analytically and numerically the response of random singlet quantum critical points to local perturbations. We showed that such systems suffer from an Anderson orthogonality catastrophe very similar to that in metals, with the overlap of the groundstates with and without a cut in the system decaying algebraically with system size. We showed that  the orthogonality catastrophe is multifractal in the sense that different powers of the disorder-averaged overlap $G$ scale with non-trivial powers of system size. We also discussed the implications for more general local perturbations and for the nonequilibrium dynamics of such systems after a local quantum quench.

Using real space renormalization group techniques, we were able to compute the corresponding orthogonality exponents exactly in some limits, and approximately for the full multifractal spectrum. Whether exact expressions can be obtained from the RSRG for the multifractal orthogonality spectrum and for the entanglement spectrum~\cite{PhysRevB.83.045110} remains an open question. We also note that the nature of the orthogonality catastrophe discussed here is quite different from that in metals: in particular, we showed that the overlap remains finite (on average) for more generic, ``softer'' local perturbations (see Sec.~\ref{subsecLocalPe} above). Similarly, coupling a random-singlet system to a quantum impurity will not lead to an orthogonality catastrophe, contrary to the metallic case. The physical mechanisms underlying the orthogonality catastrophe in random-singlet systems are also quite different from those at play in metals: within the RSRG, a physical cut in a random-singlet system leads to an avalanche of local rearrangements of singlets that in turn leads to the orthogonality catastrophe. It would be interesting to study this orthogonality catastrophe in the free fermions case (XX chain) using different approaches~\cite{PhysRevB.56.12970}, and to perhaps relate it to the more usual language of phase shifts or to the singular form of the density of states at energy $\varepsilon=0$. It would also be very interesting to investigate whether the orthogonality catastrophe discussed here generalizes to highly excited states -- in the context of many-body localization -- for systems that can be tackled using similar RSRG techniques~\cite{PekkerRSRGX}, and if it can be related to the discussion in Ref.~\onlinecite{2014arXiv1411.2616K}. 

\smallskip

{\it Acknowledgments.}
This work was supported by the Quantum Materials program of LBNL (R.V), NSF grant DMR-1206515 and the Simons Foundation (J.E.M.). We thank N. Laflorencie, A. Lazarescu, A.C. Potter, Z. Ringel, H. Saleur and M. Serbyn for insightful discussions. 

\bibliographystyle{mybibstyle}

\bibliography{MBL}

\begin{thebibliography}{10}

\bibitem{PhysRev.163.612}
G.~D. Mahan.
\newblock \emph{Excitons in metals: Infinite hole mass}.
\newblock Phys. Rev. \textbf{163}, 612--617 (1967).

\bibitem{PhysRev.178.1097}
P.~Nozi\`eres and C.~T. De~Dominicis.
\newblock \emph{Singularities in the x-ray absorption and emission of metals.
  iii. one-body theory exact solution}.
\newblock Phys. Rev. \textbf{178}, 1097--1107 (1969).

\bibitem{PhysRevLett.18.1049}
P.~W. Anderson.
\newblock \emph{Infrared catastrophe in fermi gases with local scattering
  potentials}.
\newblock Phys. Rev. Lett. \textbf{18}, 1049--1051 (1967).

\bibitem{9780511470752}
A.~C. Hewson.
\newblock \emph{The Kondo Problem to Heavy Fermions} (Cambridge University
  Press, 1993).
\newblock ISBN 9780511470752.
\newblock Cambridge Books Online.

\bibitem{Calabrese:2006}
P.~Calabrese and J.~Cardy.
\newblock \emph{{Time dependence of correlation functions following a quantum
  quench}}.
\newblock Physical Review Letters \textbf{96(13)}, 136801 (2006).

\bibitem{1742-5468-2007-10-P10004}
P.~Calabrese and J.~Cardy.
\newblock \emph{Entanglement and correlation functions following a local
  quench: a conformal field theory approach}.
\newblock Journal of Statistical Mechanics: Theory and Experiment
  \textbf{2007(10)}, P10004 (2007).

\bibitem{PhysRevLett.109.260601}
A.~Mitra.
\newblock \emph{Time evolution and dynamical phase transitions at a critical
  time in a system of one-dimensional bosons after a quantum quench}.
\newblock Phys. Rev. Lett. \textbf{109}, 260601 (2012).

\bibitem{1367-2630-12-5-055015}
D.~Fioretto and G.~Mussardo.
\newblock \emph{Quantum quenches in integrable field theories}.
\newblock New Journal of Physics \textbf{12(5)}, 055015 (2010).

\bibitem{PhysRevLett.106.227203}
P.~Calabrese, F.~H.~L. Essler and M.~Fagotti.
\newblock \emph{Quantum quench in the transverse-field {I}sing chain}.
\newblock Phys. Rev. Lett. \textbf{106}, 227203 (2011).

\bibitem{PhysRevLett.110.257203}
J.-S. Caux and F.~H.~L. Essler.
\newblock \emph{Time evolution of local observables after quenching to an
  integrable model}.
\newblock Phys. Rev. Lett. \textbf{110}, 257203 (2013).

\bibitem{PhysRevLett.98.180601}
C.~Kollath, A.~M. L\"auchli and E.~Altman.
\newblock \emph{Quench dynamics and nonequilibrium phase diagram of the
  bose-hubbard model}.
\newblock Phys. Rev. Lett. \textbf{98}, 180601 (2007).

\bibitem{PhysRevLett.98.210405}
S.~R. Manmana, S.~Wessel, R.~M. Noack and A.~Muramatsu.
\newblock \emph{Strongly correlated fermions after a quantum quench}.
\newblock Phys. Rev. Lett. \textbf{98}, 210405 (2007).

\bibitem{PhysRevB.22.1305}
C.~Dasgupta and S.-k. Ma.
\newblock \emph{Low-temperature properties of the random heisenberg
  antiferromagnetic chain}.
\newblock Phys. Rev. B \textbf{22}, 1305--1319 (1980).

\bibitem{FisherRSRG2}
D.~S. Fisher.
\newblock \emph{Random antiferromagnetic quantum spin chains}.
\newblock Phys. Rev. B \textbf{50}, 3799--3821 (1994).

\bibitem{PhysRevLett.84.3434}
K.~Damle, O.~Motrunich and D.~A. Huse.
\newblock \emph{Dynamics and transport in random antiferromagnetic spin
  chains}.
\newblock Phys. Rev. Lett. \textbf{84}, 3434--3437 (2000).

\bibitem{refaelreview}
G.~Refael and J.~E. Moore.
\newblock \emph{Criticality and entanglement in random quantum systems}.
\newblock Journal of Physics A: Mathematical and Theoretical \textbf{42},
  504010 (2009).

\bibitem{PhysRevB.83.045110}
M.~Fagotti, P.~Calabrese and J.~E. Moore.
\newblock \emph{Entanglement spectrum of random-singlet quantum critical
  points}.
\newblock Phys. Rev. B \textbf{83}, 045110 (2011).

\bibitem{white}
S.~R. White and A.~Feiguin.
\newblock \emph{Real-time evolution using the density matrix renormalization
  group}.
\newblock Phys. Rev. Lett. \textbf{93}, 076401 (2004).

\bibitem{schollwoeck}
U.~Schollw{\"o}ck.
\newblock \emph{The density-matrix renormalization group in the age of matrix
  product states}.
\newblock Annals of Physics \textbf{326(1)}, 96 -- 192.
\newblock January 2011 Special Issue (2011).

\bibitem{VoskAltmanPRL13}
R.~Vosk and E.~Altman.
\newblock \emph{Many-body localization in one dimension as a dynamical
  renormalization group fixed point}.
\newblock Phys. Rev. Lett. \textbf{110}, 067204 (2013).

\bibitem{PekkerRSRGX}
D.~Pekker, G.~Refael, E.~Altman, E.~Demler and V.~Oganesyan.
\newblock \emph{Hilbert-glass transition: New universality of temperature-tuned
  many-body dynamical quantum criticality}.
\newblock Phys. Rev. X \textbf{4}, 011052 (2014).

\bibitem{PhysRevLett.112.217204}
R.~Vosk and E.~Altman.
\newblock \emph{Dynamical quantum phase transitions in random spin chains}.
\newblock Phys. Rev. Lett. \textbf{112}, 217204 (2014).

\bibitem{HeisenbergRSRGX}
R.~Vasseur, A.~C. Potter and S.~A. Parameswaran.
\newblock \emph{Quantum criticality of hot random spin chains}.
\newblock Phys. Rev. Lett. \textbf{114}, 217201 (2015).

\bibitem{RevMBL}
R.~Nandkishore and D.~A. Huse.
\newblock \emph{Many-body localization and thermalization in quantum
  statistical mechanics}.
\newblock Annual Review of Condensed Matter Physics \textbf{6(1)}, 15--38
  (2015).

\bibitem{ReviewMBLEhud}
E.~{Altman} and R.~{Vosk}.
\newblock \emph{{Universal Dynamics and Renormalization in Many-Body-Localized
  Systems}}.
\newblock Annual Review of Condensed Matter Physics \textbf{6}, 383--409
  (2015).

\bibitem{Mirlin2000259}
A.~D. Mirlin.
\newblock \emph{Statistics of energy levels and eigenfunctions in disordered
  systems}.
\newblock Physics Reports \textbf{326(5--6)}, 259 -- 382 (2000).

\bibitem{RevModPhys.80.1355}
F.~Evers and A.~D. Mirlin.
\newblock \emph{Anderson transitions}.
\newblock Rev. Mod. Phys. \textbf{80}, 1355--1417 (2008).

\bibitem{Subramaniam2006}
A.~R. Subramaniam, I.~A. Gruzberg, A.~W.~W. Ludwig, F.~Evers, A.~Mildenberger
  and A.~D. Mirlin.
\newblock \emph{Surface criticality and multifractality at localization
  transitions}.
\newblock Phys. Rev. Lett. \textbf{96}, 126802 (2006).

\bibitem{PhysRevLett.106.107402}
H.~E. T\"ureci, M.~Hanl, M.~Claassen, A.~Weichselbaum, T.~Hecht, B.~Braunecker,
  A.~Govorov, L.~Glazman, A.~Imamoglu and J.~von Delft.
\newblock \emph{Many-body dynamics of exciton creation in a quantum dot by
  optical absorption: A quantum quench towards kondo correlations}.
\newblock Phys. Rev. Lett. \textbf{106}, 107402 (2011).

\bibitem{Latta:2011aa}
C.~Latta, F.~Haupt, M.~Hanl, A.~Weichselbaum, M.~Claassen, W.~Wuester,
  P.~Fallahi, S.~Faelt, L.~Glazman, J.~von Delft, H.~E. Tureci and A.~Imamoglu.
\newblock \emph{Quantum quench of kondo correlations in optical absorption}.
\newblock Nature \textbf{474(7353)}, 627--630 (2011).

\bibitem{PhysRevLett.108.190601}
M.~Heyl and S.~Kehrein.
\newblock \emph{Crooks relation in optical spectra: Universality in work
  distributions for weak local quenches}.
\newblock Phys. Rev. Lett. \textbf{108}, 190601 (2012).

\bibitem{PhysRevLett.110.230601}
R.~Dorner, S.~R. Clark, L.~Heaney, R.~Fazio, J.~Goold and V.~Vedral.
\newblock \emph{Extracting quantum work statistics and fluctuation theorems by
  single-qubit interferometry}.
\newblock Phys. Rev. Lett. \textbf{110}, 230601 (2013).

\bibitem{PhysRevX.2.041020}
M.~Knap, A.~Shashi, Y.~Nishida, A.~Imambekov, D.~A. Abanin and E.~Demler.
\newblock \emph{Time-dependent impurity in ultracold fermions: Orthogonality
  catastrophe and beyond}.
\newblock Phys. Rev. X \textbf{2}, 041020 (2012).

\bibitem{PhysRevLett.111.046402}
B.~D\'ora, F.~Pollmann, J.~Fort\'agh and G.~Zar\'and.
\newblock \emph{Loschmidt echo and the many-body orthogonality catastrophe in a
  qubit-coupled luttinger liquid}.
\newblock Phys. Rev. Lett. \textbf{111}, 046402 (2013).

\bibitem{PhysRevLett.110.240601}
R.~Vasseur, K.~Trinh, S.~Haas and H.~Saleur.
\newblock \emph{Crossover physics in the nonequilibrium dynamics of quenched
  quantum impurity systems}.
\newblock Phys. Rev. Lett. \textbf{110}, 240601 (2013).

\bibitem{PhysRevLett.112.246401}
M.~Schir\'o and A.~Mitra.
\newblock \emph{Transient orthogonality catastrophe in a time-dependent
  nonequilibrium environment}.
\newblock Phys. Rev. Lett. \textbf{112}, 246401 (2014).

\bibitem{PhysRevLett.112.146804}
R.~Vasseur and J.~E. Moore.
\newblock \emph{Edge physics of the quantum spin hall insulator from a quantum
  dot excited by optical absorption}.
\newblock Phys. Rev. Lett. \textbf{112}, 146804 (2014).

\bibitem{PhysRevX.4.041007}
R.~{Vasseur}, J.~P. {Dahlhaus} and J.~E. {Moore}.
\newblock \emph{Universal nonequilibrium signatures of majorana zero modes in
  quench dynamics}.
\newblock Phys. Rev. X \textbf{4}, 041007 (2014).

\bibitem{1742-5468-2011-03-L03002}
J.~Dubail and J.-M. Stephan.
\newblock \emph{Universal behavior of a bipartite fidelity at quantum
  criticality}.
\newblock Journal of Statistical Mechanics: Theory and Experiment
  \textbf{2011(03)}, L03002 (2011).

\bibitem{PhysRevB.65.081106}
Y.~Gefen, R.~Berkovits, I.~V. Lerner and B.~L. Altshuler.
\newblock \emph{Anderson orthogonality catastrophe in disordered systems}.
\newblock Phys. Rev. B \textbf{65}, 081106 (2002).

\bibitem{2014arXiv1411.2616K}
V.~Khemani, R.~Nandkishore and S.~L. Sondhi.
\newblock \emph{Nonlocal adiabatic response of a localized system to local
  manipulations}.
\newblock Nat Phys \textbf{11(7)}, 560--565 (2015).

\bibitem{FisherRSRG1}
D.~S. Fisher.
\newblock \emph{Random transverse field {I}sing spin chains}.
\newblock Phys. Rev. Lett. \textbf{69}, 534--537 (1992).

\bibitem{PhysRevB.51.6411}
D.~S. Fisher.
\newblock \emph{Critical behavior of random transverse-field {I}sing spin
  chains}.
\newblock Phys. Rev. B \textbf{51}, 6411--6461 (1995).

\bibitem{RefaelMoore}
G.~Refael and J.~E. Moore.
\newblock \emph{Entanglement entropy of random quantum critical points in one
  dimension}.
\newblock Phys. Rev. Lett. \textbf{93}, 260602 (2004).

\bibitem{PhysRevLett.99.140405}
N.~E. Bonesteel and K.~Yang.
\newblock \emph{Infinite-randomness fixed points for chains of non-abelian
  quasiparticles}.
\newblock Phys. Rev. Lett. \textbf{99}, 140405 (2007).

\bibitem{PhysRevLett.43.1434}
S.-k. Ma, C.~Dasgupta and C.-k. Hu.
\newblock \emph{Random antiferromagnetic chain}.
\newblock Phys. Rev. Lett. \textbf{43}, 1434--1437 (1979).

\bibitem{PhysRevLett.101.120603}
A.~Silva.
\newblock \emph{Statistics of the work done on a quantum critical system by
  quenching a control parameter}.
\newblock Phys. Rev. Lett. \textbf{101}, 120603 (2008).

\bibitem{1742-5468-2011-08-P08019}
J.-M. Stephan and J.~Dubail.
\newblock \emph{Local quantum quenches in critical one-dimensional systems:
  entanglement, the loschmidt echo, and light-cone effects}.
\newblock Journal of Statistical Mechanics: Theory and Experiment
  \textbf{2011(08)}, P08019 (2011).

\bibitem{PhysRevB.72.140408}
N.~Laflorencie.
\newblock \emph{Scaling of entanglement entropy in the random singlet phase}.
\newblock Phys. Rev. B \textbf{72}, 140408 (2005).

\bibitem{PhysRevB.85.094417}
F.~Igl\'oi, Z.~Szatm\'ari and Y.-C. Lin.
\newblock \emph{Entanglement entropy dynamics of disordered quantum spin
  chains}.
\newblock Phys. Rev. B \textbf{85}, 094417 (2012).

\bibitem{PhysRevLett.109.250602}
A.~Gambassi and A.~Silva.
\newblock \emph{Large deviations and universality in quantum quenches}.
\newblock Phys. Rev. Lett. \textbf{109}, 250602 (2012).

\bibitem{PhysRevB.63.134424}
O.~Motrunich, K.~Damle and D.~Huse.
\newblock \emph{Dynamics and transport in random quantum systems governed by
  strong-randomness fixed points}.
\newblock Phys. Rev. B \textbf{63}, 134424 (2001).

\bibitem{PhysRevB.56.12970}
L.~Balents and M.~P.~A. Fisher.
\newblock \emph{Delocalization transition via supersymmetry in one dimension}.
\newblock Phys. Rev. B \textbf{56}, 12970--12991 (1997).

\end{thebibliography}
    
   \end{document}